\documentclass[pra,aps,twocolumn]{revtex4}

\usepackage{epsfig}
 
\begin{document}
 
\newcommand{\capeq}[1]{Equation (\ref{eq:#1})}
\newcommand{\capsec}[1]{Section \ref{sec:#1}}
\newcommand{\eq}[1]{Eq.~(\ref{eq:#1})}
\renewcommand{\sec}[1]{Sec.~\ref{sec:#1}}
\newcommand\beq{\begin{equation}}
\newcommand\eeq{\end{equation}}
\newcommand\be{\begin{equation}}
\newcommand\ee{\end{equation}}
\newcommand\bea{\begin{eqnarray}}
\newcommand\eea{\end{eqnarray}}
\newcommand{\ket}[1]{| #1 \rangle}
\newcommand{\bra}[1]{\langle #1 |}
\newcommand{\braket}[2]{\langle #1 | #2 \rangle}
\newcommand{\proj}[1]{| #1\rangle\!\langle #1 |}
\newcommand{\ba}{\begin{array}}

\author{John A. Smolin}
 
\title{The Continuous Variable Quantum Teleportation Controversy}

\affiliation{IBM Watson Research Center, P.O. Box 218,
Yorktown Heights, New York 10598, USA}
\date{June 8, 2004}
 
\begin{abstract}
I argue that the objections by Rudolph and Sanders \cite{RS} to performing
continuous variable quantum teleportation experiments using lasers, as
well as the various rebuttals to their paper, are based on a misunderstanding 
of the Partition Ensemble Fallacy.  
\end{abstract}

\maketitle

A quantum mechanical mixed state described by a density matrix $\rho$ 
can always be decomposed into a variety of ensembles of pure states 
$\{p_i^j,\psi_i^j\}$ such that
\begin{equation}
\forall_j \sum_i p_i^j \proj{\psi_i^j} = \rho\ .
\end{equation}
Quantum mechanics tells us that no choice of ensemble is preferred over
any other.  To commit the so-called partition ensemble fallacy 
(PEF) \cite{KB} is
to retrodict a particular ensemble from one's experimental results, claiming
to have determined ``what really happened.''
This is {\em not} to say, however, that one cannot reasonably interpret 
results as having arisen from one ensemble or another.  In fact, this is
precisely what quantum mechanics allows us to do.  

A cottage industry has appeared recently disputing the claim of a
demonstration of continuous variable quantum teleportation (CVQT) by
Furusawa {\em et al.} \cite{furusawa}.  Rudolph and Sanders (RS)
assert \cite{RS} that the PEF has been committed, and that
CVQT has not in fact been demonstrated, while several
refutations of their refutation have appeared
\cite{vEF,vEF_PRL,NB,wiseman}, which mutually disagree to various
extent.  Though this author believes that many interesting, useful,
and correct results have been promulgated in this endeavor, it is the
purpose of this paper to argue that the whole controversy is
misguided, being founded upon a misapplication of the PEF.

The state of a laser is normally considered as a coherent state:
\begin{equation}
\ket{\psi}=\proj{\alpha e^{i \phi}},
\end{equation}
with a definite phase $\phi$ and a mean photon number $|\alpha|^2$.
This is not quite correct in reality, since a real laser
produces is a state 
\begin{equation}
\rho=\frac{1}{2 \pi} \int d\phi \proj{\alpha e^{i \phi}}, 
\label{rho}
\end{equation}
an equal mixture of all phases.  Experiments are typically performed
considering phase relative to a reference beam taken off at a beam
splitter and shared by all aspects of the experiment.  This is what is done
in \cite{furusawa}.  All experimental outcomes relating to the relative
phase will behave {\em as if} the state had been a pure coherent state
all along. It is this retrodiction that is cited in \cite{RS} as being 
the commission of the PEF.

This author's view is that there is no PEF to commit.  No claim needs
to be made that ``what really happened'' was that there was a coherent state,
if only we could know its phase.  Instead, the claim is about what {\em would
have happened} had a particular coherent state been used.  Quantum mechanics
tells us, and indeed we'd be committing the PEF to think otherwise, that because
we can write $\rho$ in the form (\ref{rho}), we are entitled to 
exactly this claim.  It is important to stress that this is valid 
{\em provided one believes quantum mechanics}.  Experiments in CVQT are testing
our ability to perform teleportation, not testing quantum mechanics 
itself. 
It is the failure to draw this distinction
that I would have called the ``partition ensemble fallacy fallacy,'' had
that name not already been taken \cite{NB}, so I will refer to it instead
as the partition ensemble fallacy fallacy type 2 (PEFF2).


Let us now examine RS's objections, as well as the various responses to
RS in light of this viewpoint:

\subsection{Rudolph and Sanders \protect\cite{RS}}
As RS point out, since $\rho$ (\ref{rho}) is a mixed state, it is valid
to consider other ensembles leading to the same density matrix, and we
can write $\rho$ in terms of number states using the identity 
\begin{equation}
\rho=\frac{1}{2 \pi} \int d\phi \proj{\alpha e^{i \phi}}=
\sum_n e^{-|\alpha|^2} \frac{|\alpha|^{2n}}{n!}\proj{n}\ .
\label{identity}
\end{equation}
In this decomposition, diagonal in the number state basis, no teleportation
can occur, since $\rho$ {\em has} no well defined phase at all.  RS go
so far to avoid the PEF that they do not give this ensemble a more special 
place than (\ref{rho}); their point is that no decomposition is privileged.
As I have already argued, I consider this a non issue according to the
PEFF2.

RS have several more specific complaints, related to the three criteria
for successful CVQT \cite{BFK}.

\noindent $\bullet$ a) The states to be teleported should be unknown
to Alice and Bob (AB), and supplied by ``an actual'' third party
``Victor.''

Since in the actual experiment Victor shares a reference laser with AB, 
his state is not truly independent.  Written in the number basis
the resulting joint density matrix shows significant correlation between
AB and Victor.  This is hardly a surprise as the point of
sharing the reference laser with Victor was so his phase would be
correlated with AB's.  On the other hand, working in the
basis of coherent states it is simple to see that if victor inserts a phase
shift into his beam, AB can have no information about the
shift, even if their overall phases are correlated with Victor's.  Appealing
to the PEFF2 confirms that this is acceptable.

Still, the spirit of quantum teleportation is for Victor to share
nothing with AB, save the state he hands to Alice to
be teleported, and the reference laser violates that spirit.  A
better experiment would have Victor use his own independent laser, though
this should not in principle change the results.  It is not clear
that everyone agrees with this point, and we shall return to it by and by.

\noindent $\bullet$ b) Alice and Bob share only a nonlocal entangled resource and
a classical channel through which Alice transmits her measurement
results to Bob.

Just as Victor gets to share in the reference laser, in the
experimental setup of \cite{furusawa} Alice and Bob share the
reference beam during the entire course of the teleportation.  This in
itself violates criterion b (it is something extra other than
entanglement or a classical channel), but worse, it implies they in
fact share a quantum channel during the teleportation.  Perhaps
quantum information is being sent through the channel channel,
cheating on the experiment of performing true teleportation.  Wiseman,
as well as van Enk and Fuchs (vEF) \cite{wiseman,vEF,vEF_PRL}, point out
that this could be addressed by sharing the time reference before the
teleportation commences, just as the entanglement must be shared
first, ensuring quantum information can not be surreptitiously passed
from Alice to Bob during the protocol.  This is an additional shared
resource without which teleportation cannot be done, but we must live
with this.  Teleportation as originally conceived
\cite{teleportation} assumes such a resource--a shared basis is
required even for simple qubit based protocols.

This presharing of the reference laser has not been done in
actual experiments, and this author believes it should be before a
complete CVQT experiment can be claimed.  

\noindent $\bullet$ c) Entanglement should be a verifiable resource.
RS show that the density matrix of the squeezed state generated 
starting from the mixed state $\rho$ (\ref{rho}), rather than
from a pure coherent state, is separable.  But it is demonstrated
in \cite{vEF,vEF_PRL} that when one includes the reference beam 
and considers the entire state of AB that they do indeed
share {\em distillible} entanglement.  

It is not obvious that merely sharing distillible entanglement without
actually distilling it into pure entanglement is sufficient for
teleportation.  However, it is shown \cite{furusawa} that the
fidelity achieved in the experiment exceeds that which could be
achieved by a classical measure and resend scheme.  The fact that this
calculation was performed using the coherent state ensemble merely
reduces the argument back to the PEF and PEFF2.  In a way,
criterion c is not really a separate criterion at all.  No one really
cares if entanglement is used, only that the fidelity achieved with
entanglement exceeds what could be achieved without it.  Checking for
entanglement is simply one way to establish whether the protocol is
classical or not.  This leads us to the following discussion:

\subsection{The privileged role of Victor}

In a teleportation protocol Victor, the verifier, plays a special
role.  It is he who gets to decide if AB have successfully
teleported the state he asks them to.   Consider a situation where
he as inserted AB's teleporter into one arm of an interferometer
and where he can vary the phase in that arm in a way unknown to Alice and Bob.
If Victor uses a pure coherent state and the fidelity
at the output is high, he will consider AB to have teleported
his state correctly.  He does not care at all whether AB
really had entanglement, or about any of their problems with the PEF.
If, on the other hand, Victor also has to use the same sort of mixed
states as (\ref{rho}), he can still check the fidelity AB
achieve in their teleportation apparatus, but he is left scratching
his own head over the PEF. This leaves us with several questions:

\noindent $\bullet$ 
1) If Victor has pure coherent states, but AB have to use 
states of the form (\ref{rho}), can they teleport Victor's states with 
high fidelity?

Even if AB have a phase relative to Victor's
pure state that is unknown to them, the goal of teleportation is to
faithfully transfer unknown states.  Including an additional unknown
phase shift is of no consequence.  

A teleportation protocol for $S$ for AB acts on an input pure state $\ket{\psi_\phi}$ with
phase $\phi$ and leads to an output density matrix $S(\psi_\phi)$ with the property
that the fidelity 
is high for all input states:
\begin{equation}
\forall_\phi  \bra{\psi_\phi}S(\ket{\psi_\phi})\ket{\psi_\phi}) \ge f_{\rm min}
\label{fidelity}
\end{equation}
Introducing an unknown phase $\eta$ between Victor and AB means $S$ acts on a phase-shifted
state and leads to a density matrix for Victor
\begin{equation}
\rho_V = 
\frac{1}{2\pi} \int\! d\eta\, \Phi_\eta S(\ket{\psi_{\phi+\eta}}) \Phi_\eta^\dag \ ,
\end{equation}
where $\Phi_\eta \ket{\psi_{\phi+\eta}} \equiv \ket{\psi_\phi}$.  Then, using (\ref{fidelity})
\begin{eqnarray}
\nonumber &\rho_V=
\frac{1}{2\pi} \int\! d\eta\, \Phi_\eta [f_{\rm min} \proj{\psi_{\phi+\eta}}+(1\!-\!f_{\rm min}) \rho_{\phi\eta} ]\Phi_\eta^\dag \\
&=\frac{1}{2\pi} \int\! d\eta [f_{\rm min} \proj{\psi_\phi}+ (1\!-\!f_{\rm min}) \Phi_\eta\rho_{\phi\eta}\Phi_\eta^\dag]
\end{eqnarray}
where $\rho_{\phi\eta}$ is a density matrix which in general might depend on $\phi$ and $\eta$.  Finally, since the
second term is positive, $\rho_V$ has fidelity $F(\rho_V,\ket{\psi_\phi})\ge f_{\rm min}$.

\noindent $\bullet$ 
2) If Victor also must use states of the form (\ref{rho}), can he
really know if AB could have teleported a pure coherent state,
and can we really say they have achieved teleportation?
The answer is yes, by the now tiresome appeal to the PEFF2.

\noindent $\bullet$ 3) In the real experiments \cite{furusawa} Victor
shares the reference beam. Would the experiment still work if Victor
had his own independent laser, and can we really say we have tested
teleportation?  This question is the acid test for having
achieved teleportation, and surprisingly it does not appear to be
crisply answered in most of the literature \cite{vEFnote}.  Perhaps it
is considered such an obvious consequence as to be not worth
mentioning.

Answering question 3 is a combination of the first two answers.  An additional
laser with another unknown phase will not change AB's ability
to teleport a phase unknown to them.  And Victor's test of their ability
is valid under the PEFF2.  This view is apparently not accepted by 
all, some not mentioning the issue and vEF making the odd claim
that Victor can use his own laser, but only if he phase locks it
with AB's reference laser. They tell us that ``Alice's 
claim is only that she can teleport a quantum state of a particular mode:
Victor is free to choose the state to be teleported, but not the Hilbert 
Space.''  To me this suggests that Victor must have a laser of the same
frequency (the same mode) as AB's, but the phase is exactly
the variable being teleported, and therefore must be free.

Because of this disagreement and, as I have mentioned earlier, having Victor
share the reference with AB goes against the spirit of
teleportation, it would seem the experiment with Victor having his
own laser ought to be performed.

\subsection{\label{vEFsection} van Enk and Fuchs \protect\cite{vEF,vEF_PRL}}

In this paper vEF show that the state of an ideal 
propagating laser field divided into packets of some duration $T$
can be thought of as the tensor product of coherent state packets
of unknown phase, but all sharing the same unknown phase.  They then
use the quantum de Finetti Theorem \cite{df1,df2} to show that
this is the only valid tensor-product decomposition of the corresponding
density matrix.  This is a compelling suggestion that the decomposition
(\ref{rho}) plays a privileged role, at least when describing propagating
lasers.  It is also an extremely useful formulation, making quite clear, for
example, how to calculate the fidelity of a CVQT experiment,
and why there is distillible entanglement in AB's (mixed) squeezed
state.

Rudolph and Sanders do not find this convincing, however \cite{RS1}, 
reiterating the point that $\rho$ remains a mixed state and no 
decomposition is truly privileged.  They argue that while the vEF
formulation makes some things easy to understand if one prefers
the decomposition (\ref{rho}) but that there is still no necessary reason
to prefer it.  I again suggest the entire debate is ill posed.

\subsection{Wiseman \protect\cite{wiseman}}
In \cite{wiseman}, Wiseman argues extensively against the existence of
an absolute phase and for the idea that a laser is as good a clock to
use for an agreed time or phase standard as any other.  His
dismiss both the claim by RS \cite{RS} that
continuous-variable quantum teleportation has not been achieved (one
way to state their claim is that a laser rather than an absolute clock 
was used to synchronize Alice and Bob) and the refutation of RS 
by vEF \cite{vEF} (who still implicitly appeal to the idea of absolute 
phase).  
Wiseman says a laser is the best thing one can use as a time standard; 
{\em the PEFF2 says that it doesn't matter if a there is or is not a better 
time standard}.  These points of view are, in a sense, two sides of the same
coin and if Wiseman were to argue that they are indeed the same,
I would not strongly disagree--the discussion will have long since moved
closer to philosophy than physics.

I doubt than anyone would seriously disagree with Wiseman's contention that
there is no better clock at optical frequencies than a laser, and that
the best we can do is measure phase relative to a laser as standard.  
What {\em is} easy to object to is the idea 
of going
around willy-nilly defining laser phases as 0 relative to themselves.
When Alice and Bob have one laser, and Victor has another, they cannot
both be defined as 0 phase (though this is OK provided they never
do anything to compare phase, nor does anyone perform a calculation
depending on the phase difference).

Wiseman states ``[his own] arguments lead inevitably to the conclusion
that in quantum optical experiments there is no necessity to consider,
even hypothetically, any time-keeper beyond the laser which serves as
a phase reference.  No other clock is superior in any fundamental
sense.'' and that ``it is precisely because no experiment is affected
by the supposed randomness in the phase of the laser (if it is being
used as a time reference) that makes it possible to describe the laser
by a {\em single} state ....''

The trouble with this, of course, is in deciding who is right when there 
is more than one phase reference laser.  When AB purport
to teleport Victor's state, Victor is well justified in saying his
clock is superior to theirs in the fundamental sense of ``the customer
is always right.''  From his point of view (relative to {\em his} clock,
AB's reference laser {\em is} in a mixed state, and they
cannot claim otherwise.  As I have argued, teleportation 
works regardless, but there certainly are experiments where AB's 
randomness of phase affects the outcome.  Victor's own randomness
of phase relative to the implicit but arguably unmeasurable absolute phase 
is unimportant for the reasons given by Wiseman, as well as
for those given here.  The advantage in the PEFF2 formulation
over Wiseman's is that it still makes sense whether or not a reference
laser is shared by all.

A secondary point concerns Wiseman's discussion of criterion b.  He
agrees that the experiment could be performed with Alice's and Bob's
clocks synchronized before the teleportation commences, avoiding all
chance of quantum information sneaking though the clock
synchronization channel.  He goes on to argue 
that a classical channel could be used to synchronize clocks
\cite{dephase}, so that even if it were being used during the
teleportation, no quantum information could sneak through.  But then
he makes a logical mistake.

Wiseman claims that since time synchronization channel could be
dephased and the experiment would still work, that this is just as
good as performing the experiment with a dephased channel and that the
onus is on the debunker to dephase the channel if he wants to disprove
the experiment.  But this is wrong.  The onus is always on the one who
claims to have performed and experiment to
convince the skeptics not the other way around.
It is not enough to claim that the experiment would have worked,
if only we hadn't cheated.  This does sound dangerously like commission of the 
PEFF2, which allows us to make claims about ``what would have
happened.''  The difference is that in one case we are arguing 
whether an apparatus has been shown to teleport; 
in the other the argument is over whether the apparatus should
be called a teleporter at all.

Wiseman also wants to claim ``by fiat'' that the synchronization laser
can be considered classical because other classical systems (marbles are
his example) are really quantum mechanical and only called classical
by fiat in some approximation.
But marbles typically have wavefunctions like 
$\psi_0=\frac{1}{\sqrt{2 \pi \sigma^2}} e^{-(x-x_0)^2/(2\sigma^2)}$.
If one were to create a superposition of these states 
$\frac{1}{\sqrt{2}}(\ket{\psi_0}+\ket{\psi_1})$ it would almost immediately
collapse to $\ket{\psi_{0,1}}$.  This natural decoherence
due to the environment is why marbles are classical, it is not
merely an arbitrary choice.  Coherent states of lasers do not
suffer the same decoherence, or at least do so only at a vastly
different timescale.  Indeed if they did, we wouldn't be able
to even consider using them for CVQT.

\subsection{The Partition Ensemble Fallacy Fallacy Fallacy}

Nemoto and Braunstein state (NB) \cite{NB} that the choice of a flat
distribution over phases in (\ref{rho}) is unfounded, and that any
choice of distribution is in fact unfalsifiable.  Since choosing a
single phase is also unfalsifiable, we (they argue) may do so thus
invalidating the PEF and RS's argument as applied to CVQT.  This is a
quite similar point of view to that of Wiseman \cite{wiseman}, and has
the same flaw.  NB call this the ``partition
ensemble fallacy fallacy'' (PEFF).  The PEFF is itself fallacious, or
at least NB's application of it is.

Their mistake is in believing that just because absolute phase is
unobservable (itself contentious) that the distribution over
phases of a laser state is also unobservable.  If one were to measure
the phases of many different lasers relative to a single reference
laser, one could experimentally determine the distribution of their
phases, even without knowing the absolute phase of the reference.  In
other words, imagine Victor measuring AB's phase relative
to his own--we expect it will be random with the distribution
(\ref{rho}).  

Because this distribution over phases is measurable, there is no ``freedom
of religion'' over choice of distribution as there is over choice
of ensembles that lead to the same density matrix.  The PEFF2 does not
have this same weakness--it explains why laser CVQT experiments are acceptable independently
of whether either absolute phase or the distribution over phases are
falsifiable.

\subsection{Conclusions}

It is useful to consider whether regular qubit based teleportation protocols
share the same conceptual difficulties as CVQT.  The analogous situation
is on in which Alice, Bob, and Victor each have a qubit bases modified
by a random unitary operators each time they perform the experiment.
The reason their bases must be randomized each time is to ensure
they employ states of the form $\rho^{\otimes n}$ rather than
$\int d\psi (\proj{\psi})^{\otimes n}.$  This is in analogy to a laser
having a new random phase each time it is turned on (and for real lasers, any
time past their coherence time).

For their teleportation to succeed, Alice and Bob will have to coordinate bases
before attempting to send Victor's state.  Victor, on the other hand, need
not contrive to align his basis with some outside observer's basis (be
it some fourth party Carmela, or a preferred basis of the universe).
I don't think anyone would question that Victor is testing AB's
teleportation experiment despite the fact that when he creates a pure state
to give to Alice, Carmela considers it a completely mixed state.
If Victor chooses to think of his state as a pure state, that convenient
fiction is well justified by the PEFF2.


To sum up:  I have argued that CVQT is possible with traditional 
laser sources, despite their lack of absolute phase.  The actual
experiments are slightly unsatisfactory in that Alice and Bob share
the reference laser during the teleportation, and Victor shares 
the reference rather than having his own laser beam, but neither
of these weaknesses are fundamental.

The objections to these experiments appealing to the PEF are based
on a misunderstanding of the PEF that we are retrodicting ``what really
happened'' rather than {\em predicting} what would given a particular
state to teleport.

Some of the objections to RS's objections are based on a (appropriate)
denial of the reality of absolute phase, but become confused (or at
least confusing) by presence of more than one laser in the world,
not all of which can simultaneously be defined as 0 phase. 
It is also confusing that teleportation is a protocol that
always works, and indeed is designed to always work, on inputs of
any relative phase to Alice and Bob.  Because of this their
absolute phase and phase relative to anyone aside from each
other is unimportant.  This fact tends to obscure
the already subtle situation of phase in CVQT experiments.

\noindent {\bf Acknowledgments}:  The author thanks G. Burkard and
I. Devetak for helpful discussions, and the US National
Security Agency and the Advanced Research and Development Activity 
for support through contract DAAD19-01-C-0056.


\begin{references}
\bibitem{vEFnote} vEF do address this question in \protect\cite{vEF,vEF_PRL},
though their conclusion is confusing.
\bibitem{dephase} By ``classical channel'' Wiseman means a completely dephasing 
channel in some basis.  This is called
in some contexts a quantum-classical-quantum channel.
\end{references}
\end{document}